# Electric field driven octahedral rotation in perovskite

Wonshik Kyung[1,2,3,‡], Choong H. Kim[1,2,‡], Yeong Kwan Kim[4], Beomyoung Kim[3], Chul Kim[5], Woobin Jung[1,2], Junyoung Kwon[1,2], Minsoo Kim[1,2], Aaron Bostwick[3], Jonathan D. Denlinger[3], Yoshiyuki Yoshida[6], and Changyoung Kim[1,2,*]

[1] Center for Correlated Electron Systems, Institute for Basic Science (IBS), Seoul 08826, Republic of Korea

[2] Department of Physics and Astronomy, Seoul National University (SNU), Seoul 08826, Republic of Korea

[3] Advanced Light Source, Lawrence Berkeley National Laboratory, California 94720, USA

[4] Department of Physics, KAIST, Daejeon 34141, Republic of Korea

[5] Institute of Physics and Applied Physics, Yonsei University, Seoul 03722, Korea

[6] National Institute of Advanced Industrial Science and Technology, Tsukuba 305-8568, Japan

Email address of corresponding author: changyoung@snu.ac.kr

**Abstract**

Rotation of $MO_6$ (M = transition metal) octahedra is a key determinant of the physical properties of perovskite materials. Therefore, tuning physical properties, one of the most important goals in condensed matter research, may be accomplished by controlling octahedral rotation (OR). In this study, it is demonstrated that OR can be driven by an electric field in $Sr_2RuO_4$. Rotated octahedra in the surface layer of $Sr_2RuO_4$ are restored to the unrotated bulk structure upon dosing the surface with K. Theoretical investigation shows that OR in $Sr_2RuO_4$ originates from the surface electric field, which can be tuned via the screening effect of the overlaid K layer. This work establishes not only that variation in the OR angle can be induced by an electric field, but also provides a way to control OR, which is an important step towards in situ control of the physical properties of perovskite oxides.

**Introduction**

Perovskite materials possess some of the most interesting properties in condensed matter physics, such as superconductivity, metal-insulator transitions and ferroicity.[1] Theoretical and experimental research has proven that octahedral rotation (OR) plays an important role in those properties. For instance, OR significantly affects metal-insulator transitions[2,3] and exotic orbital-selective phenomena[4,5] by changing the inter-site electron hopping probability and even the structural symmetry of materials. Furthermore, the magnetic ground state of a material is often governed by OR through altered super-exchange or Dzyaloshinskii-Moriya interactions.[6] Therefore, the ability to readily vary OR would be an important step towards controlling such physical properties.

However, in spite of its importance, there are a limited number of reports on controlling OR.[7-9] The main reason for this is that the OR angle is thought to be an inherent characteristic of a material, determined by the steric effect arising from the sizes of its constituent atoms.[10,11] Thus, most of the attempts to control OR have been limited to substitution of atoms with different ionic sizes or application of strains using different substrates. However, these methods do not truly allow control of the OR angle as they cannot be applied in situ, and typically cause complex side effects.[12,13]

When attempting to control physical properties through the OR angle, various parameters should be considered, such as pressure and magnetic and electric fields. Among these parameters, the electric field has distinct advantages in terms of convenience, controllability and minimal power consumption. In this letter, we report electric-field dependent evolution of the OR angle on the surface of $Sr_2RuO_4$. By means of in situ dosing with potassium (K), the surface electric field can be tuned through the screening effect of the overlying K atoms. $Sr_2RuO_4$ was chosen as the target material given its distinct surface layer-driven bands, which arise due to differences in its structural symmetry (a finite OR angle at its surface and no rotation of its bulk);[14-16] these properties not only provide clean surfaces, but also make investigation of its electronic structure relatively easy. Our results obtained using the surface-sensitive techniques of low-energy electron diffraction (LEED) and angle-resolved photoelectron spectroscopy (ARPES) indicate that reduction of the electric field results in a reduced OR angle (down to zero). Our density functional theory (DFT) studies have shown that the surface electric potential (or surface electric field)[17,18] is responsible for the OR on the surface of $Sr_2RuO_4$, which in turn implies the possibility of controlling the OR angle via an electric field.

**Results**

**Disappearance of $Sr_2RuO_4$ surface states upon K coverage**

Figure 1 shows Fermi surface (FS) maps of freshly cleaved and K-dosed $Sr_2RuO_4$ surfaces. The main features are two bulk electron pockets ($\beta^b$ and $\gamma^b$) centered at $\Gamma$ and a bulk hole pocket ($\alpha^b$) centered at ($\pi,\pi$). In addition to these bulk FSs, we detected additional FSs from the surface ($\alpha^s$, $\beta^s$, $\gamma^s$) and zone-folded surface ($\alpha^{sf}$, $\beta^{sf}$, $\gamma^{sf}$) bands (Figure 1a), consistent with previously reported ARPES results.[15,16,19] Notably, when we deposit K on the surface with coverage above one monolayer (ML), all of the surface-related FSs disappear (Figure 1b), which is somewhat similar to the case of surface aging of $Sr_2RuO_4$.[16,19] To deduce the reason for this disappearance, we systematically varied the coverage of K while conducting LEED and ARPES measurements.

LEED is one of the most direct and convenient methods to investigate in-plane crystal symmetry. In case when there is no OR distortion, only the 1 × 1 peak should appear. When the in-plane OR angle finite, the unit cell is transformed to a $(\sqrt{2} \times \sqrt{2})R45°$ unit cell, and consequently additional fractional peaks ($\sqrt{2} \times \sqrt{2}$) should appear in the LEED pattern. The relative intensity of the fractional to integer spot is approximately proportional to the OR angle.[20,21] The LEED results from our systematic investigation of K coverage are shown in

Figure 2a. In the LEED data of pristine Sr$_2$RuO$_4$, we can observe both 1 × 1 and $\sqrt{2} \times \sqrt{2}$ peaks. As the K coverage increases, the OR-driven $\sqrt{2} \times \sqrt{2}$ peaks gradually weaken and eventually disappear above 1 ML of K, while the 1 × 1 peaks remain robust. Therefore, we suspect that the K layer gradually and eventually completely suppresses the OR of the surface Sr$_2$RuO$_4$.

The ARPES results from our systematic investigation of K coverage are shown in Figure 2b; they also provide evidence for the suppression of OR in the surface Sr$_2$RuO$_4$. The data for pristine Sr$_2$RuO$_4$ along Γ-M-Γ (Figure 2b) show $β^b$, $β^s$, $γ^b$ and $γ^s$ bands. As the K coverage increases, the $β^s$ band, instead of being suppressed, moves toward $β^b$ before finally merging with it. As for the γ bands, $γ^b$ remains the same while the spectral weight of the $γ^s$ band at the M-point gradually weakens. These observations indicate that the difference between the bulk and surface electronic structures gradually decreased, consistent with the suppression of OR in the surface layer. Our detailed analysis regarding the ARPES spectra (Supplementary Figure 1) also supports that the suppression of the surface layer-driven bands is induced by suppression of the surface OR.

**Cause of the suppressed octahedral rotation**

The next question is why the OR angle reduces with K-dosing. Surface alkali metal atoms can play roles in electron doping, chemical bonding and changing the surface electric potential; we shall consider each in turn. First, we consider the possibility that the vanishing OR is caused by electron transfer from the alkali metal to the Sr$_2$RuO$_4$. To investigate this, we obtained the electron occupations of the bands from their FS volume; we list these in the Supplementary Table 1 along with all other values discussed here. We find that there is an occupation difference of 0.09 electrons between the fresh surface (3.82) and that with 1 ML of K (3.91), which agrees well with the number (0.07) extracted from our theoretical study (Supplementary Figure 2). However, an average transfer of 0.09 electrons from the K atoms is not likely to cause complete suppression of OR considering the case of (Sr,La)$_2$RuO$_4$, which is regarded as electron-doped Sr$_2$RuO$_4$ (Supplementary Figure 3). On the other hand, effects arising from chemical bonding cannot provide an adequate explanation either; our measured K 3$p$ core-level spectra as a function of K coverage (Supplementary Figure 4) clearly show the absence of chemical bonding associated with K atoms. Furthermore, as mentioned, our experimental and theoretical results (Supplementary Table 1 and Supplementary Figure 2) suggest that there an electron transfer of only 0.07-0.09 from the K atoms, which is too small to be due to chemical bonding between the K and O atoms.

As the change in OR angle cannot be attributed to electron doping or chemical bonding effects, let us now consider the gradient of the surface electric potential. In general, the electric potential is modulated due to the electric potential difference between vacuum and solid. This electric potential modulation is greatly affected by the surface condition, as illustrated in Figures 3a and 3b. To investigate how electric potential relates to the OR angle, we performed a DFT calculation which has been widely used to describe low energy physics in Sr$_2$RuO$_4$,[15,16,22] by constructing a five-layer slab of Sr$_2$RuO$_4$ with and without an ML of K overlaid. We also performed calculations for various K layer distances away from the equilibrium position, to investigate the evolution of the electric potential and OR angle.

First, we note that the potential energy decreases gradually towards the value of the interlayer region as the K layer approaches the equilibrium position (the arrow in Figure 3b). This suggests that the role of the K layer is to mitigate the surface electric field; 1 ML of K causes the electric potential in the surface region to become similar to the interlayer potential. Consequently, the surface layer is in a bulk-like potential, which should lead to the suppression of OR (Figure 3a). The calculated OR angle monotonically evolves from zero to fully rotated as the K layer distance increases from the equilibrium position (Figure 3c). These observations strongly indicate that the origin of the OR is in the electric field, consistent with the experimental data in Figure 2.

Even though it is not essential to prove the electric origin of the OR, it is noteworthy that the OR evolution in the calculation in Figure 3c fairly consistently reproduces the behavior according to the thickness of the K layer seen in the experimental data. Partial K-coverage cases have been simulated by artificially moving the K layer away from the equilibrium position.[23] The similarity between the experimental and theoretical results may be understood in the following way. Figure 3b shows the calculated electric potential as a function of the K layer distance from surface layer. The electric potential at the interface (between the outermost SrO layer and the K layer) monotonically and gradually changes as the K layer distance changes. We expect that the surface electric field will be screened by K atoms proportionally with the K coverage (Figure 3a). Therefore, the 'moving K layer distance' should generate similar trends in the electric potential as does the K coverage. Our electronic structure with 'moving K layer distance' does indeed show trends consistent with our K coverage dependent ARPES results (Supplementary Figure 5); hence, the method should be reasonable to observe the overall trend.

To directly investigate the role of the electric field in the OR angle, we performed another five-layer slab calculation, this time with an external electric field applied perpendicular to the surface (Figure 3d). Our DFT calculations of total energy predict that the surface OR changes, exhibiting behavior proportional to the increase of ~ 0.15° per 0.1 V $Å^{-1}$ (Figure 3d); thus, the electric field appears to be coupled to the OR. Therefore, we conclude that the electric field is responsible for the OR in the $Sr_2RuO_4$ surface layer and thus the OR can be varied by tuning the electric potential.

**Mechanism of the electric field driven octahedral rotation**

The next step is to find out how an electric field couples with the OR. Previous theoretical studies have shown that ferroelectric-like atomic displacement competes with the OR, which is the reason why most of ferroelectric materials do not have OR distortion.[24] Therefore, it is natural to consider non-uniform atomic displacements driven by depth-dependent electric field as the cause for the OR angle change. From our DFT calculation with K-layer distance variation, we extract atomic displacement in the outermost Sr-O layer. It shows that the distance between upper and lower Sr atoms in the surface $Sr_2RuO_4$ layer is the most sensitive factor to the K-layer distance (Supplementary Figure 6). It is found that the vertical Sr-Sr distance (defined in Figure 3a) increases with the electric field (K-layer distance) and changes more than 0.1 Å as shown in Figure 4a

In order to check whether the (vertical) Sr-Sr distance is coupled to the OR angle variation, we have estimated OR angle from bulk $Sr_2RuO_4$ calculation as a function of the Sr-Sr distance (Figure 4b). As can be seen in Figure 4b, variation of the Sr-Sr distance successfully reproduces emergence of the OR angle in $Sr_2RuO_4$ in the range over which the Sr-Sr distance varies in Figure 4a. Therefore, it appears that the Sr-Sr distance is the mediating parameter between the surface electric field and OR angle. As the stronger (weaker) surface electric field makes a larger (smaller) Sr-Sr distance, and it eventually leads to a larger (smaller) OR angle. In this context, we can reconsider how OR varies depending on the situation in $Sr_2RuO_4$. Absence of surface electric potential makes octahedron unrotated in bulk $Sr_2RuO_4$. On the other hand, the surface layer feels the surface electric potential, and thus the (surface electric potential driven) large Sr-Sr distance makes octahedron rotated in the surface layer. Upon K dosing, the K atoms gradually reduce the surface electric potential toward the value of bulk electric potential. Then, the OR angle in the surface layer gradually decreases down to zero as we observed.

This mechanism can be applied to various materials other than $Sr_2RuO_4$. We performed additional bulk calculation on $Sr_2RhO_4$ and $Sr_2IrO_4$ for various Sr-Sr distances (Figure 4b), and it is found that OR angles of $Sr_2RhO_4$ and $Sr_2IrO_4$ are affected by Sr-Sr distance in the same manner as in the case of $Sr_2RuO_4$, even though the large initial OR angles of $Sr_2RhO_4$ and $Sr_2IrO_4$ make their variation less dramatic. These results not only consistently simulate the OR angle variation behavior in K dosed $Sr_2RuO_4$, but also suggest that there might be universal coupling between the OR angle and cation distance in TMO.

**Discussion**

We shall now discuss the difference between the local and external electric fields to allow our result to be more fully understood. Even though we theoretically demonstrated an OR angle change by an external electric field (Figure 3d), ~ 4 V Å$^{-1}$ is required for full suppression of the OR, which is not practical. However, a local electric field (potential gradient near the surface) can induce full suppression of the OR in Sr$_2$RuO$_4$, as shown in Figure 3b, where varying the K layer distance changes the electric potential energy on the order of a few eV. This case demonstrates the advantages of exploiting the local electric field, in terms of strength and controllability. In practice, such local electric fields could be achieved by using special techniques such as ionic gating or even exploiting existing, naturally occurring fields at interfaces. In that respect, issues in groups of material heterostructures, which exhibit numerous exotic phenomena that the corresponding bulk crystals do not, such as superconductivity,[25,26] metal-insulator transitions[27] and controllable ferromagnetism could be revisited.[28] We suspect that local electric fields at interfaces play a significant role in generating these phenomena, and our findings should provide important clues regarding their microscopic mechanisms.

Since OR is not a polar distortion, whereas an electric field would produce a polar distortion, an electric field effect has rarely been considered when determining the OR angle. In this regard, the discovery of hybrid improper ferroelectricity (HIF) was a surprise because it shows an unexpected coupling between OR and ferroelectric (polar) distortion of Ca atom (A-atom in A$_3$B$_2$O$_7$) in double-layer perovskites such as Ca$_3$Ti$_2$O$_7$.[29,30] Some theoretical studies have proposed electrical control of physical properties by changing the OR angle via the HIF mechanism,[31,32] but this has not yet been achieved experimentally. Contrary to the case of HIF, Sr$_2$RuO$_4$ is a metal and normally is not a suitable candidate for ferroelectric distortion. However, a finite displacement for Sr atom (A-atom in A$_2$BO$_4$) in the surface Sr$_2$RuO$_4$ layer can be induced by the surface electric potential, and then the Sr displacement may establish the connection between electric field and OR (Figure 4). Note that A-atom mediates electric field and OR in both HIF and surface $Sr_2RuO_4$ cases, which may imply a possible universal mechanism of OR angle variation via A-atom modulation. Our work not only is to show a change OR angle via electric field, but also may initiate follow-up studies to elucidate the mechanism of OR angle variation.

Finally, let us discuss possible effects of OR angle variation in systems with strong electron correlation since perovskite oxides are typically in the strong electron correlation regime. Since OR angle can significantly affect the exchange interaction as well as electron localization, it is expected that change in the OR angle can lead to various phenomena, e.g., magnetism and metal-insulator transitions. Such effects may be found in the case of $Ca_{2-x}Sr_xRuO_4$ (CSRO) which may be viewed as $Sr_2RuO_4$ with OR angle variation. In CSRO, rich and complex phases,[33-35] such as magnetisms (ferromagnetism[33] or antiferromagnetism[34]), appear and emergence of those phenomena are attributed to OR distortions.[5,36,37] Another important aspect is that the electronic structure of $Sr_2RuO_4$ possesses a van Hove singularity near the Fermi level whose position is very sensitive to the OR angle. The large density of states from the van Hove singularity can boost the instability from electron correlation. This implies that we could effectively control the electron correlation strength of the system via the OR angle. Control of OR therefore may allow us a more diverse controllability of physical properties in strongly correlated materials.

In conclusion, our experimental and theoretical investigations demonstrate that variation of electric potential is responsible for the OR angle in the $Sr_2RuO_4$ surface layer, and that OR angle can be varied by tuning the electric potential through surface K dosing. Our result not only sheds light on the mechanism of octahedral distortion found in various oxide systems but is also an important step towards electric field control of physical properties via variation of the OR angle in perovskite oxides.

**Methods**

**ARPES and LEED measurement conditions**

ARPES ($hv$ = 70 eV) measurements were performed at beam lines (BLs) 4.0.3 and 7.0.2 of the Advanced Light Source, Lawrence Berkeley National Laboratory, USA. Potassium dosing was carried out by evaporating K onto the sample using commercial alkali metal dispensers (SAES). Spectra were acquired with R8000 (BL 4.0.3) and R4000 (BL 7.0.2) electron analyzers (Scienta). Total energy resolution was set to 12 meV, and the angular resolution was 0.00163 Å$^{-1}$. Cleaving and dosing of the samples were done at 20 K in an ultra-high vacuum better than $5 \times 10^{-11}$ Torr. LEED measurements were performed at BL 7.0.2 of the Advanced Light Source and at the Center for Correlated Electron Systems of the Institute for Basic Science, Seoul National University, Republic of Korea using a LEED spectrometer (SPECS) with an electron energy of 187 eV.

**Information of DFT calculation**

As the feasibility of using DFT to understand the structural and electronic properties of $Sr_2RuO_4$ has been shown through previous studies,[15,16,22] we performed first-principle calculation using the non-spin-polarized DFT method without spin-orbit coupling. The Perdew-Burke-Ernzerhof form of the exchange-correlation functional was used as implemented in VASP software.[38,39] We used a 600 eV plane wave cut-off energy and 12 × 12 × 1 $k$-points for all calculations and the projector augmented wave method. The in-plane lattice constant was fixed at the experimental value of $Sr_2RuO_4$. All the internal atomic positions were fully relaxed until the maximum force was below 0.5 meV Å$^{-1}$ while the symmetries of the system (point group D$_{4h}$) were maintained during the relaxation. In practice, since full relaxation is numerically unstable, we fixed the rotation angle of the octahedron and relaxed only the vertical positions of the atoms of the surface. In this way, the energy curve as a function rotation angle was obtained, and the angle with the energy minimum was found. We also checked that no additional symmetry lowering occurs even without symmetry constraints for a few cases. To mimic partial K coverage, we performed a five-layer slab calculation with 15 Å vacuum layer, which is symmetric with respect to the middle layer with an overlying K atom layer. In this calculation, we relaxed the distance between the K layer and the $Sr_2RuO_4$ to find the equilibrium K layer position. The resultant value was the 'K EQ position' in Figure 3 (9.4 Å$^{-1}$ from the second Ru-O layer). In the calculation, we relaxed both the outermost layer and the K layer distance, which explains why we defined the position relative to the second Ru-O layer. The location of each K atom could not be specified experimentally, so we assumed it to be above the apical oxygen atom of the outermost Sr-O layer (as illustrated in Figure 3a) since that is the energetically most stable position in our DFT calculation. For the bulk calculations of $Sr_2RhO_4$ and $Sr_2IrO_4$, the in-plane lattice constants were also fixed as the experimental lattice constant of each compound, 5.4516 and 5.4956 Å, respectively. In these calculations, we fixed the Sr and Ir positions and allowed to move oxygen atoms only to find the optimum rotation angle of the octahedron for a given Sr-Sr distance.

**Data Availability**

The datasets generated during and/or analyzed during the current study are available from the corresponding author upon request.

**Acknowledgements**

We are grateful to E. A. Kim and S. Y. Park for fruitful discussions. This work was supported by the Institute for Basic Science in Korea (Grant No. IBS-R009-G2). The Advanced Light

Source is supported by the Office of Basic Energy Sciences of the U.S. DOE under Contract No. DE-AC02-05CH11231.

**Author Contributions**

‡W. K. and C. H. K contributed equally to the work. W. K. conceived the work. W. K., Y. K. K., B. K. and C. K. (Chul Kim) performed ARPES measurements with the support from J. D. D., and W. K. and Y. K. K. analyzed the data. W. K., W. J., J. K., and M. K. performed LEED measurement with the support from A. B. Samples were grown and characterized by Y. Y.. C. H. K. led the theoretical study and performed DFT calculation. All authors discussed the results. W. K. and C. K. (Changyoung Kim) led the project and manuscript preparation with contributions from all authors.

**Competing interests**

The Authors declare no Competing Financial or Non-Financial Interests.

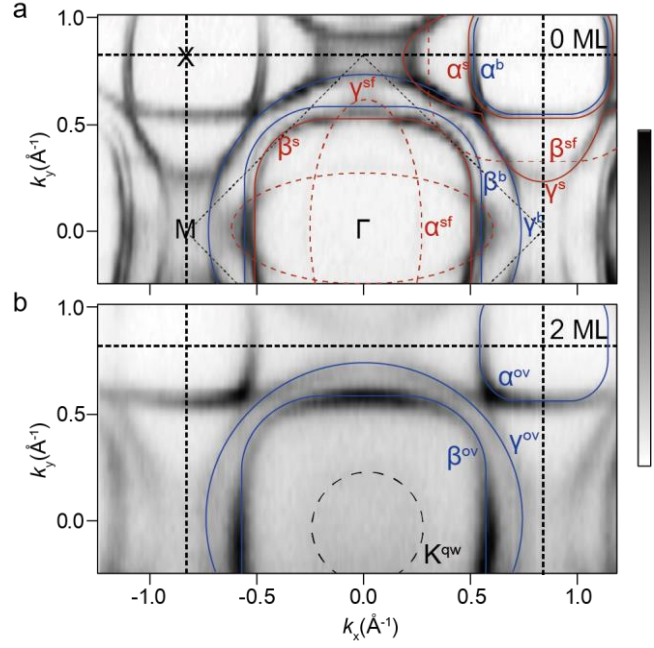

**Figure 1. Fermi surface (FS) maps of Sr$_2$RuO$_4$.** a) Fresh Sr$_2$RuO$_4$ and (b) Sr$_2$RuO$_4$ covered in two monolayers (MLs) of K. The black thick (thin) dashed lines mark the Brillouin zone of the bulk (surface) Sr$_2$RuO$_4$ without (with) octahedral rotation. RuO$_6$ octahedra on the surface rotate, resulting in a $\sqrt{2} \times \sqrt{2}$ reconstructed surface that causes replica FSs to appear. Superscripts b, s, sf and ov denote bulk, surface, surface folding and overlap, respectively. The blue (red) guidelines indicate the bulk (surface) bands. K$^{qw}$ indicates a K circular quantum well state (Supplementary Figure 4).

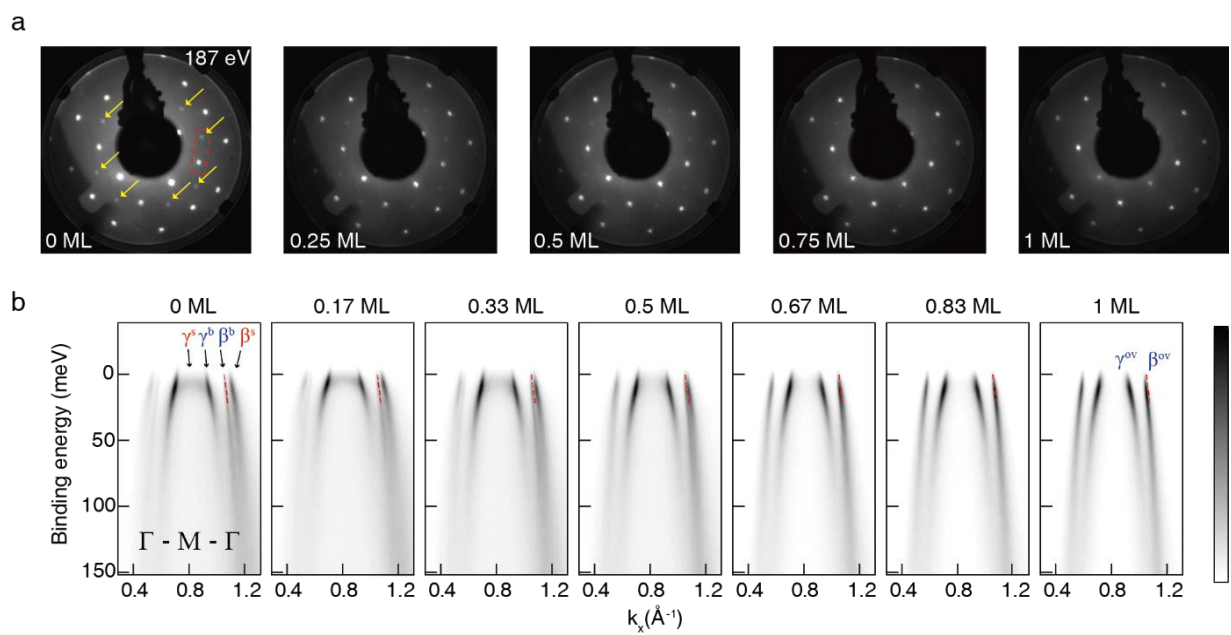

**Figure 2. Low-energy electron diffraction and angle-resolved photoelectron spectroscopy results according to the K coverage.** a) Electron diffraction images for various K coverages (0, 0.25, 0.5, 0.75 and 1 ML). The yellow arrows indicate peaks due to $\sqrt{2} \times \sqrt{2}$ surface reconstruction; as the K coverage increases, these peaks gradually become weaker and eventually disappear. The rectangle in red is the region of quantitative intensity analysis in Supplementary Figure 1. b) Photoelectron spectroscopy images along Γ–M–Γ for various K coverages (0, 0.17, 0.34, 0.5, 0.67, 0.83 and 1 ML). Red dashed lines are to mark the position of $\beta^b$ of 0 ML coverage.

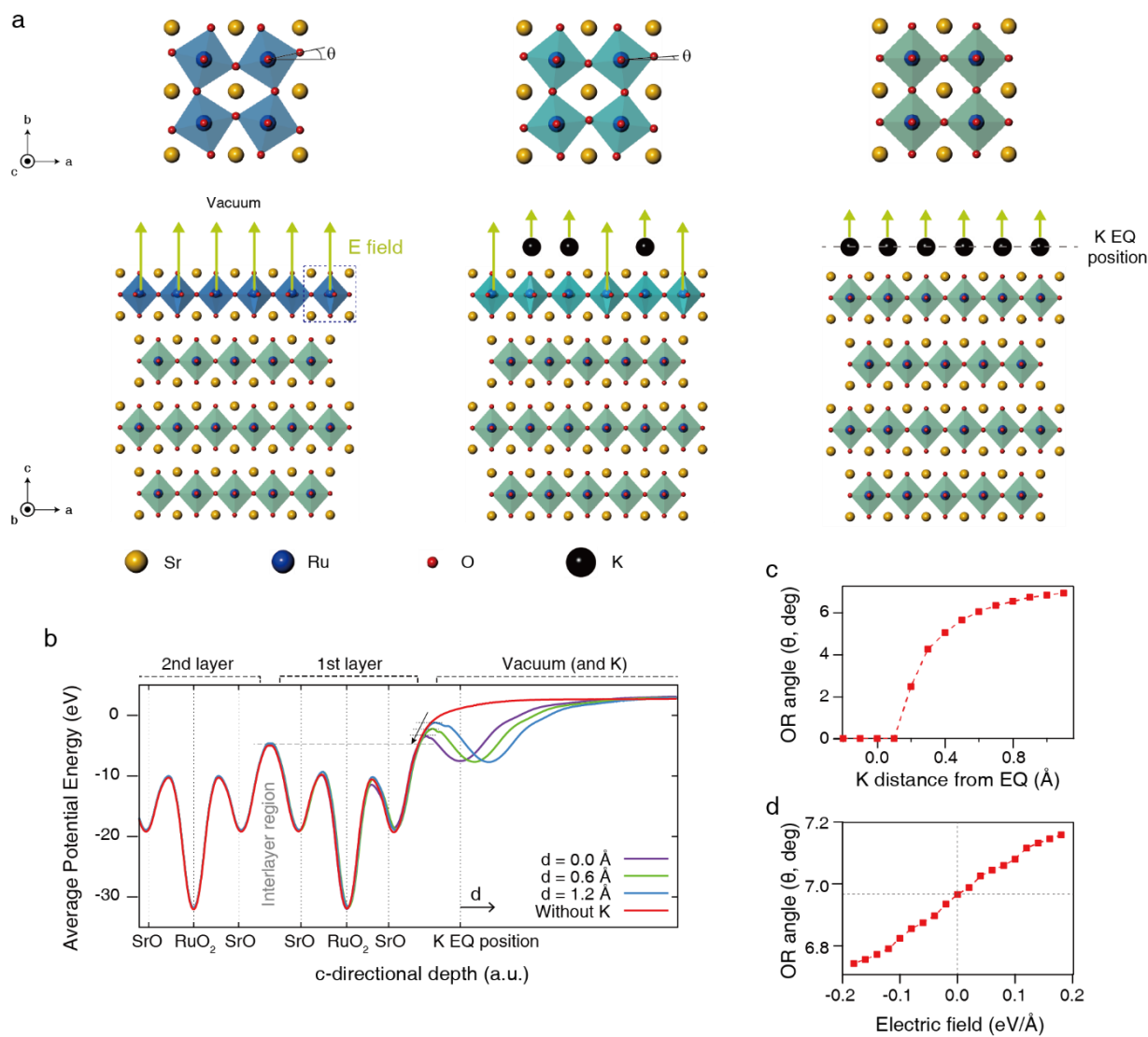

**Figure 3. Electric potential driven octahedral rotation (OR).** a) Crystal structure of $Sr_2RuO_4$ with three different K coverages (left: fresh, middle: partial coverage, right: 1 ML). The dashed rectangle marks the area for the enlarged view in Figure 4a. b-d) Density functional theory (DFT) results for a slab of $Sr_2RuO_4$ five layers thick. b) Electric potential energy as a function of the depth in the c-direction (surface normal), for various K layer distances from the equilibrium position: $d$ = 0.0, 0.6, 1.2 Å and without a K layer. The equilibrium position of the surface K layer (9.4 Å from the second outermost Ru-O layer) was obtained from the DFT calculation. The grey horizontal dashed line indicates the electric potential energy in the $Sr_2RuO_4$ interlayer region. As the K layer distance decreases, the electric potential in the SrO-K region gradually decreases toward the value of the interlayer region (black arrow). c) Plot of OR angle as a function of the K layer distance from the equilibrium position. d) Plot of OR angle as a function of electric field.

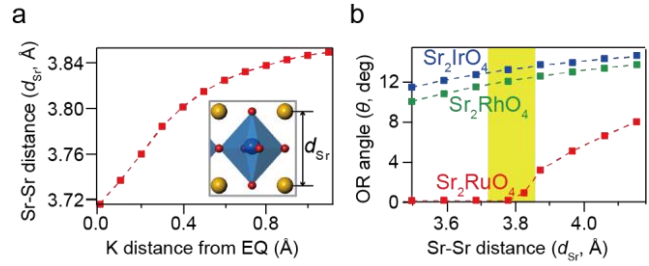

**Figure 4. Vertical Sr-Sr distance as the link between the surface electric potential and OR angle.** a) DFT calculation results from a five-$Sr_2RuO_4$-layer slab. The plot shows the Sr-Sr distance as a function of the K layer distance. Inset: enlarged view of an octahedron in the surface $Sr_2RuO_4$ layer. Sr-Sr distance ($d_{Sr}$) is defined as the vertical distance between upper and lower Sr atoms. b) DFT results for the OR angle as a function of the Sr-Sr distance for bulk $Sr_2RuO_4$, $Sr_2RhO_4$ and $Sr_2IrO_4$. The shaded region marks the actual range over which the Sr-Sr distance varies in a).

# Supplementary information of 'Electric field driven octahedral rotation in perovskite'

Wonshik Kyung[1,2,3,‡], Choong H. Kim[1,2,‡], Yeong Kwan Kim[4], Beomyoung Kim[3], Chul Kim[5], Woobin Jung[1,2], Junyoung Kwon[1,2], Minsoo Kim[1,2], Aaron Bostwick[3], Jonathan D. Denlinger[3], Yoshiyuki Yoshida[6], and Changyoung Kim[1,2,*]

[1] Center for Correlated Electron Systems, Institute for Basic Science (IBS), Seoul 08826, Republic of Korea

[2] Department of Physics and Astronomy, Seoul National University (SNU), Seoul 08826, Republic of Korea

[3] Advanced Light Source, Lawrence Berkeley National Laboratory, California 94720, USA

[4] Department of Physics, KAIST, Daejeon 34141, Republic of Korea

[5] Institute of Physics and Applied Physics, Yonsei University, Seoul 03722, Korea

[6] National Institute of Advanced Industrial Science and Technology, Tsukuba 305-8568, Japan

**Supplementary Note 1. Electronic structure evolution as a function of K coverage**

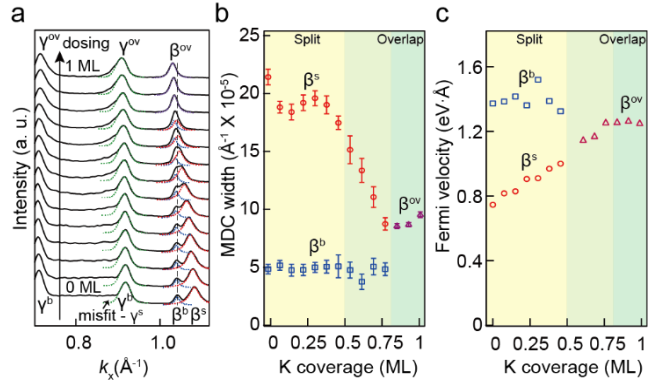

**Supplementary Figure 1:** ARPES line shape analysis as a function of K coverage. a) Momentum distribution curves (MDCs) at the Fermi energy along the Γ-M-Γ cut as a function of K coverage. Colored dashed lines are Lorentzian fits (Green: $\gamma^b$, Blue: $\beta^b$, Red: $\beta^s$), and black dashed line marks the position of $\beta^b$ of 0ML of K. The deviation on the left side of the $\gamma^b$ peak comes from $\gamma^s$ which diminishes when K coverage increases. b, c) Dosing dependent (b) MDC widths and (c) Fermi velocities of $\beta^s$ and $\beta^b$ bands.

To further prove that the suppression of surface states is not from degradation but from suppression of the surface OR, we analyze the line shape of experimental ARPES spectra in Figure 2b in the main text for more information. Since the β Fermi surface has larger difference between the surface and bulk components than the α band and is easier to track the change than the γ band, we focus on quantitative analysis of the β band line shape. Supplementary Figure 1a shows momentum distribution curves (MDCs) along the Γ-M-Γ cut at the Fermi energy as a function of K coverage, extracted from Figure 2b. As the K coverage increases, the $\beta^s$ band moves closer to the $\beta^b$ and eventually merges with it. The merging of the $\beta^s$ with the $\beta^b$ band upon K dosing, instead of disappearing, is a strong indication for suppression of the OR angle to zero (similar to bulk). On the way to merging of $\beta^s$ and $\beta^b$ bands, we observe an infinitesimal shift of the $\beta^b$ band as well (above 0.7 ML), and we believe that it comes from the second surface layer, which weakly feels the broken symmetry from the outermost layer.[1]

In order to get more information, we analyzed the momentum distribution curve (MDC) widths by fitting the $\beta^s$ and $\beta^b$ with Lorentzian functions (Supplementary Figure 1b). If the disappearance of the surface band is related to degradation of the surface layer, it is natural to expect broadening of MDCs with increasing K coverage. However, our data show that the MDC width of the surface band decreases to the value of bulk band, which is again an indication for restored OR in the surface layer. On the other hand, in comparing the $\beta^s$ and $\beta^b$ Fermi velocities, we find that the Fermi velocity of the surface band is smaller than that of the bulk but converges to the bulk value when K coverage increases (Supplementary Figure 1c). This can be attributed to the fact that the reduced Ru-O-Ru hopping under octahedra rotation recovers its bulk value when the rotation angle vanishes. Therefore, considering all of our experimental observations (disappearance of $\sqrt{2} \times \sqrt{2}$ peaks in LEED, merge of surface and bulk bands, reduction in MDC width, and enhancement of the Fermi velocity), we conclude that the disappearance of surface states upon K dosing is from reduction of OR angle to zero.

**Supplementary Note 2. Experimentally determined electron occupation for each orbital**

|   | Bulk | Surface | K dosed (2ML) |
|---|---|---|---|
| α | 1.77 | 1.73 | 1.79 |
| β | 0.86 | 0.71 | 0.87 |
| γ | 1.21 | 1.38 | 1.25 |
| Sum | 3.84 | 3.82 | 3.91 |
| (α + β)/2 | 1.31 | 1.22 | 1.33 |

**Supplementary Table 1.** Electron occupation estimated from the Fermi surface volume for bulk and surface bands of pristine sample as well as merged bands after K-dosing. The (α + β)/2 value represents the average electron occupation for $d_{yz}$ (or $d_{zx}$) orbital.

In order to figure out how the electron occupation of each orbital changes, we extracted values from Fermi surface volumes of our experimental data (Supplementary Table 1). Interestingly, difference in the total electron occupation for surface (3.82) and K-dosed (2ML) (3.91) cases is 0.09, a much smaller value than what is expected from fully ionized K. However, it is consistent with the value from calculation (Supplementary Figure 2)
The carrier densities for the pristine surface, bulk and K-dosed (2ML) are 3.82, 3.84 and 3.91, respectively. We note that the values for the pristine surface and bulk deviate from the ideal value of 4 expected for $Sr_2RuO_4$. There may be several reasons why such deviation occurs. One possibility comes from Sr vacancies generated during the cleaving of $Sr_2RuO_4$ in the vacuum to get clean surface. According to a previous report on this issue,[2] Sr deficiency can be generated when we cleave $Sr_2RuO_4$ in vacuum. Since Sr is an electron donor, Sr deficiency should result in a carrier number less than 4. However, considering carrier density in bulk (3.84) extracted from our ARPES data, Sr vacancies originated from cleavage cannot fully explain the missing carrier densities in bulk.

## Supplementary Note 3. Calculated electron occupation for each orbital

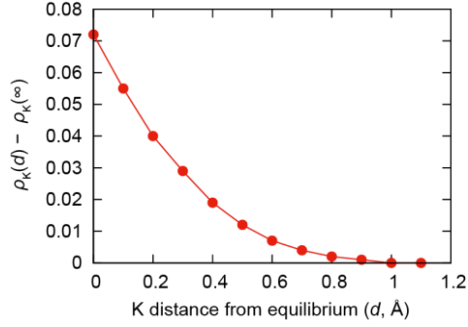

**Supplementary Figure 2.** Charge number from K layer to $Sr_2RuO_4$ as a function of the K-layer distance from the equilibrium position ($d$). The maximum electron number which transferred from K layer to $Sr_2RuO_4$ is 0.07e.

In order to estimate the amount of the electron doping by the K layer, we calculate the amount of electron at K atom as a function of K-layer distance from the equilibrium position ($\rho_K(d)$) for 1ML of K. The difference between $\rho_K(d)$ and $\rho_K(\infty)$ gives the electron transfer from the K layer to $Sr_2RuO_4$ as shown in Supplementary Figure 2. Our result shows that the electron transfer is only 0.073 electrons even for $d = 0$. This value is not too far away from the value extracted from our ARPES results (Supplementary Table 1, 0.09 electrons). That is, electrons from K atoms mostly stay around K atoms and are not transferred to $Sr_2RuO_4$.

To check whether the amount of electron transfer from K layer in $Sr_2RuO_4$ is significantly less than other cases, we also checked out several other cases based on published data.[3-5] We find 0.1 electrons for Co doped $BaFe_2As_2$,[3] 0.1-0.12 electrons for FeSe,[4] and 0.17 electrons for $YBa_2Cu_3O_7$.[5] One can see that electron transfer for $Sr_2RuO_4$ is comparable to that for Fe-based superconductors. This is surprising because the electron doping effect can be easily seen in the case of Fe-based superconductors but is hardly noticeable for $Sr_2RuO_4$. The difference comes from the difference in the Fermi surface volume; while Fe-based superconductors have relatively small Fermi surfaces, $Sr_2RuO_4$ has large Fermi surfaces.

Therefore, with a similar doping amount, the effect can be seen more easily for Fe-based superconductors.

As for the variation in the electron transfer amount among different materials, we guess that it may be related to the work function variation among different materials.

**Supplementary Note 4. Role of electron doping in octahedral rotation**

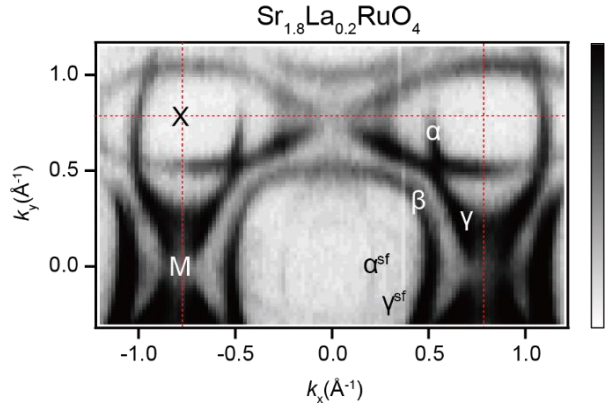

**Supplementary Figure 3.** Fermi surface map of $Sr_{1.8}La_{0.2}RuO_4$. There are folded bands of $\alpha^{sf}$ and $\gamma^{sf}$ pockets, indicating finite octahedral rotation angle for the surface layer.

The role of charge carrier doping in OR has not been understood well. In order to check whether electron doping can possibly affect the OR angle, we investigate the electronic structure of $Sr_{1.8}La_{0.2}RuO_4$. According to previous studies on $(Sr,La)_2RuO_4$, the main effect from substitution of Sr by La in $(Sr,La)_2RuO_4$ is electron doping to $RuO_2$ plane, with a robust structural symmetry (I4/mmm) up to 0.27 electron doping.[6] On the other hand, our ARPES result from $Sr_{1.8}La_{0.2}RuO_4$ (Supplementary Figure 3) shows that electron doping, up to 0.2 electrons, from La substitution cannot suppress the surface layer OR (robust $\alpha^{sf}$ and $\gamma^{sf}$ pockets). This observation suggests that, in spite of the additional 0.2 electrons, octahedra in the surface layer of $Sr_{1.8}La_{0.2}RuO_4$ are rotated as in $Sr_2RuO_4$. This in turn proves that doping of 0.09 electrons in the K dosing on $Sr_2RuO_4$ is not the reason for the completely suppressed OR.

**Supplementary Note 5. K dosing effects and their comparison with effects of aging method.**

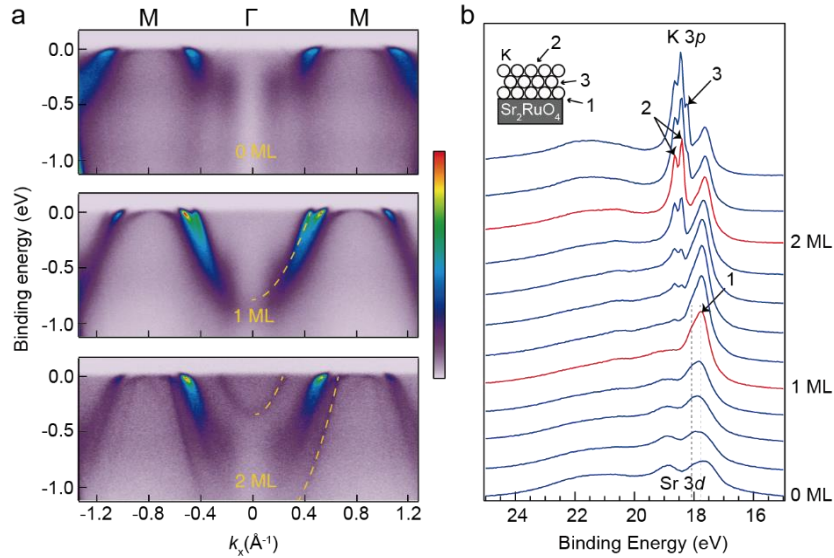

**Supplementary Figure 4.** K coverage dependent ARPES images and K 3*p* core level spectra. a) ARPES data along the high symmetry line Γ-M-Γ for 0, 1 and 2 ML of K. Yellow dashed lines indicate the K layer driven quantum well states. b) K 3*p* core level spectra as a function of K coverage. The numbers 1, 2 and 3 indicate K 3*p* peaks corresponding to K-$Sr_2RuO_4$ interface, surface and bulk K as obtained in previous studies. Dashed lines indicate the position of two spin-orbit components, K $3p_{3/2}$ and K $3p_{1/2}$, in K-$Sr_2RuO_4$ interface peak.

We calibrated the K dosing rate by monitoring the quantum well states and K 3*p* core level spectra. The ARPES data in Supplementary Figure 4a shows well-defined free-electron-like quantum well states of the K-layer, which indicate that K atoms form a well-defined two-dimensional layer with very little interaction with $Sr_2RuO_4$. The number of quantum well bands depends on the K coverage. Meanwhile, Supplementary Figure 4b shows K coverage dependent K 3*p* core level spectra. In general, different types of chemical bonding states result in core peaks at different binding energies. It is seen from the spectra in Supplementary

Figure 4b that the initial deposition of K results in the first set of peaks from #1 layer (see the inset), sitting on top of the spin-orbit split Sr $3d$ peaks. As the K coverage increases above 1 ML, we also have peaks from the surface K atoms (#2) and sandwiched atoms (#3) as previously identified.[7] Looking at the spectra, there are no peaks other than the non-bonding spin-orbit pair K $3p$ peaks, already known from previous K dosing studies.[7-9] These observations indicate absence of chemical bonding between K atoms and $Sr_2RuO_4$ as well as K intercalation effect.

At this stage, it is also worth comparing the K dosing method with previously used aging method.[1,10] The aging method usually refers to a situation when foreign atoms/molecules settling on the surface of target materials. It may lead to physisorption and/or chemical bonding, depending on what the target sample and foreign atoms/molecules are. When only physisorption is considered, the effect of aging method should be similar to the that of K dosing: reduction of surface electric potential. In this case, the only difference between aging and K dosing is whether a systematic investigation is possible in a controlled way. Otherwise, if chemical bonding effect is dominant, the chemical bonding should destroy the homogeneity of the surface. This is just a degradation of the surface, and thus it is not relevant to our 'surface potential driven octahedral rotation mechanism' case.

**Supplementary Note 6. Calculated band structure for various K layer distances**

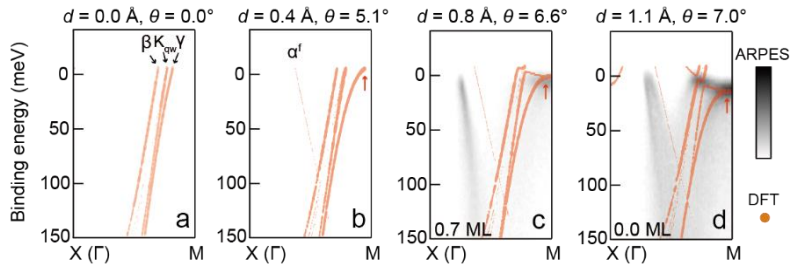

**Supplementary Figure 5.** Calculated band structures along the Γ-M direction as a function of the K-layer distance with comparison to ARPES data. a-d) Band structure of $Sr_2RuO_4$ surface bands when the K-layer distance from the equilibrium position (EQ position) $d$ is (a) 0, (b) 0.4, (c) 0.8 and (d) 1.1 Å. The red arrow indicates the surface γ band which shows a significant change as the K layer distance increases. ARPES data taken along the X-M (X corresponds to Γ in the reduced BZ) direction for K (c) 0.7 ML and (d) 0.0 ML, respectively, are overlaid. As the K coverage increases, the γ band shifts upward, which is consistent with our DFT result. $K_{qw}$ in (a) indicates K quantum well state. The DFT results are unfolded and weight of the surface band are expressed as the size of orange circles.

The justification for the 'moving K-layer distance' method to mimic the partial K coverage in experiments is given in main text. However, reliability of the method may be checked further through comparison of the calculated band structure with the experimental (ARPES) result. In order to do it, we performed band structure calculation for various K layer distances as shown in Supplementary Figure 5. We also have additional ARPES data taken along the X (Γ) - M direction and overlaid them in Supplementary Figures 5(c-d) (K 0.7 ML and 0.0 ML, respectively). Here, we unfolded the original zone-folded band structure to make it simpler and thus easy to understand.

Most significantly, the γ band (marked by the black arrow) changes from a steep electron band to a flattened hole band as the K layer distance increases, forming the well-known van Hov singularity.[11] Our ARPES data in Supplementary Figures 5(c-d) show that increase in the K coverage makes the flattened hole band (surface γ band) move upward, which is consistent with the trend in Supplementary Figure 5. That is, the 'moving K-layer distance' method used in our DFT calculation reproduces the experimental ARPES data very well, making our DFT results reliable.

**Supplementary Note 7. Effect of octahedron elongation on the octahedral rotation.**

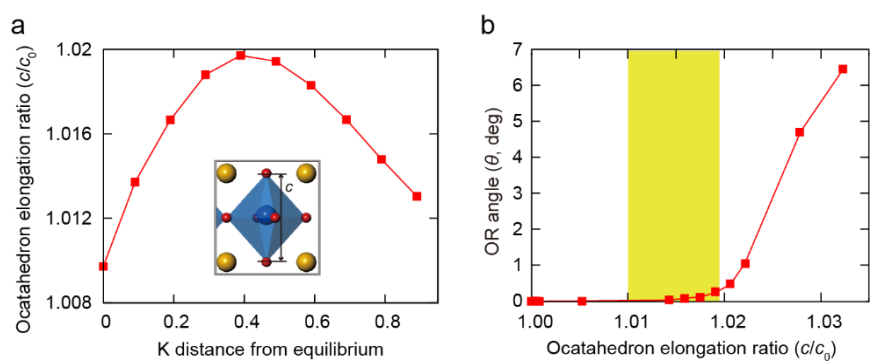

**Supplementary Figure 6.** Effect of octahedron elongation along the c-axis on OR. a) DFT calculation results from a five-$Sr_2RuO_4$-layer slab. The plot shows the octahedron elongation ratio ($c/c_0$) as a function of the K layer distance, where $c$ and $c_0$ are the vertical distance between the upper and lower apical oxygen in the surface layer and bulk, respectively. b) Calculated OR angle as a function of $c/c_0$ for the bulk. The shaded region marks the range over which the $c/c_0$ varies in a).

In an effort to find other possible links between electric potential and OR angle, we also investigated whether displacement of apical oxygen atoms in the outermost Sr-O layer is relevant to OR angle. As can be seen in Supplementary Figure 6a, the extracted $c/c_0$ from our DFT calculation does not show a monotonic behavior with the K layer distance. Meanwhile, our experimental results suggest that the OR angle is gradually reduced with the K coverage. Therefore, the non-monotonic behavior of $c/c_0$ with the K layer distance is not consistent with our experimental results.

In addition, we performed bulk DFT calculation with varying $c/c_0$ to check how OR angle varies as a function of $c/c_0$ (Supplementary Figure 6b). The results show that octahedron elongation effect in the relevant $c/c_0$ range (shaded region in Supplementary Figure 6b) is extremely small that one can conclude $c/c_0$ is not a crucial parameter for OR in $Sr_2RuO_4$.

In brief, considering nonmonotonic variation and the range of $c/c_0$, $c/c_0$ is not the mediating parameter between electric potential and OR angle in $Sr_2RuO_4$.